\documentclass{iaus}
\usepackage{graphics}
\newcommand{\pdt}[1]{{{\partial #1}\over {\partial t}}}

\newcommand{\Mp}{M_p}

\newcommand{\kapa}{\kappa_A\left(T,P\right)}
\newcommand{\kapp}{\kappa_P\left(T,P\right)}
\newcommand{\kapr}{\kappa_R\left(T,P\right)}
\newcommand{\er}{E_{R}}
\newcommand{\bbody}{B\left(T\right)}

\title[JD 11.~~Radiative Hydrodynamical Studies of Irradiated Atmospheres] 
{Radiative Hydrodynamical Studies of Irradiated Atmospheres}

\author[Ian Dobbs-Dixon]   
       {Ian Dobbs-Dixon$^1$}

\affiliation{$^1$CITA National Fellow, McGill University, Montreal,
Canada \\ email: {\tt iandd@physics.mcgill.ca}}

\pubyear{2008}
\volume{253}  
\pagerange{1--7}
\jname{IAU 253: Transiting Planets}
\editors{Queloz D., Sasselov D., Torres M., \& Holman M., eds.}
\begin{document}

\maketitle

\begin{abstract}
Transiting planets provide a unique opportunity to study the
atmospheres of extra-solar planets. Radiative hydrodynamical models of
the atmosphere provide a crucial link between the physical
characteristics of the atmosphere and the observed properties. Here I
present results from 3D simulations which solve the full Navier-Stokes
equations coupled to a flux-limited diffusion treatment of radiation
transfer for planets with 1, 3, and 7 day periods. Variations in
opacity amongst models leads to a variation in the temperature
differential across the planet, while atmospheric dynamics becomes
much more variable at longer orbital periods. I also present 3D
radiative simulations illustrating the importance of distinguishing
between optical and infrared opacities.

\end{abstract}

\firstsection 
\section{Introduction}
The abundance of close-in planets coupled with the wealth of
information that can be deduced from the fraction that transit their
host stars is now allowing us to probe our understanding of the
physics of atmospheres through radiative hydrodynamical modelling. A
number of approaches to this problem exist within the community. The
physics of radiation and hydrodynamics are both extremely important,
and there are several approaches currently utilized. Below I separate
the approaches to both the dynamical and radiation aspects. Although
this is not an exhaustive list, I believe it encapsulates the most
widely utilized approaches in the field.\\
\\
{\bf Dynamical Methods}\\ 
\indent Equivalent Barotropic Equations (2D)\\
\indent Shallow Water Equations (3D) \\
\indent Full Navier-Stokes Equations (3D)\\
{\bf Radiative Transfer Methods}\\
\indent Relaxation Methods \\
\indent Flux-limited Radiative Diffusion \\
\indent Coupled Radiation/Thermal Energy Equations \\
\indent Full Wavelength Dependent Radiative Transfer \\
\\
Ideally, we would solve the full 3D Navier-Stokes coupled with
wavelength dependent radiative transfer. However, such a task is well
beyond our computational resources, so groups working in this area
select a method from one or both lists with which to tackle the
problem. Although selecting methods with many built-in assumptions
allows for high resolution studies, one runs the risk of missing
fundamental physics crucial for a full understanding of the
atmospheres. On the flip side, as with all numerical studies, the more
exact your method, the more difficult and longer it takes to
simulate. For a comparison of methods please see \cite[Dobbs-Dixon \&
Lin (2008)]{dobbsdixon2008} or \cite[Showman et al (2008)]{showman2008}.

In the following proceedings I will describe results from simulations
utilizing the full 3D Navier-Stokes equations (described in Section
(2)). In Section (3), I couple these with the flux-limited radiative
transfer in simulations which solve a single equation for the energy
of the radiation and gas components. These simulations exhibit a range
of behavior for different opacities and orbital periods. I show that
the local value of opacity is crucial in determining the efficiency of
heat re-distribution to the night-side. In Section (4) I present
results of simulations again solving the full Navier-Stokes equations,
but now solving separate energy equations for the radiation and gas
components. Decoupling these equations allows for the introduction of
multiple opacities corresponding to both incident stellar and local
radiation. As I show, this distinction is crucial in reproducing the
observed upper atmosphere temperature inversions. I conclude in
Section (5) with a discussion of the results.

\section{Navier-Stokes Equations}
Planetary atmospheres are simulated using a three-dimensional radiative
hydrodynamical model in spherical coordinates
$\left(r,\phi,\theta\right)$. The equations of continuity and momentum
are given by
\begin{equation}
\pdt{\rho} + \nabla\cdot\left(\rho{\bf u}\right) = 0
\label{eq:continuity}
\end{equation}
and
\begin{equation}
\pdt{{\bf u}} + \left({\bf u}\cdot\nabla\right) {\bf u}= -
\frac{1}{\rho}\nabla{P} + \frac{1}{\rho}\nabla{\Phi} -2{\bf
\Omega\times u} - {\bf\Omega\times}\left({\bf\Omega\times r}\right)
\label{eq:momentum}
\end{equation}
where $\Omega$ is the rotation frequency and the $\Phi=-{G\Mp\over r}$
is the gravitational potential. 

As mentioned above, I present two separate sets of simulations in this
proceeding. For those presented in Section (3), the radiative and
thermal portions are combined into a single equation. The energy
equation can then be written as
\begin{equation}
\left[ \pdt{\epsilon} + ({\bf u}\cdot\nabla) \epsilon \right] = - P \,
\nabla \cdot {\bf u} - \nabla \cdot {\bf F} 
\label{eq:thermalenergy}
\end{equation}
where $\epsilon=c_{v} \rho T$ is the internal energy density, $T$ is
the local gas temperature, $c_v$ is the specific heat, and ${\bf F}$
is the radiative flux. The numerical techniques utilized in solving
Equations (\ref{eq:continuity})-(\ref{eq:thermalenergy}) are described
in \cite[Dobbs-Dixon \& Lin (2008)]{dobbsdixon2008}.

The final ingredient for solving the energy balance in the atmosphere
is a closure relation linking the flux ${\bf F}$ back to the radiation
energy density. Here we utilize the flux-limited diffusion (FLD)
approximation of \cite[Levermore \& Pomraning (1981)]{levermore1981},
where
\begin {equation}
{\bf F} = - \lambda {\frac{c}{\rho\kapr}} \nabla\er.
\label{eq:flux}
\end {equation}
In the above $\er=aT^4$ and $\lambda$ is a spatial and temporally
variable flux-limiter providing the closure relationship between flux
and radiation energy density. The functional form of $\lambda$ is
given by
\begin{equation}
\lambda = \frac{2+R}{6+3R+R^2},
\label{eq:lambda}
\end {equation}
where
\begin {equation}
R = \frac{1}{\rho \kapr} \frac{| \nabla\er |}{\er}.
\label{eq:Rdef}
\end {equation}
FLD allows for the simultaneous study of optically thin and optically
thick gas, correctly reproducing the limiting behavior of the radiation
at both extremes. In the optically thick limit the flux becomes the
standard
\begin {equation}
{\bf F} = - \frac{4acT^3}{3\rho\kapr} \nabla T.
\end {equation}
For the optically thin streaming limit, Equation (\ref{eq:flux})
becomes $|{\bf F}| = c\er$. Between these limits Equation (\ref{eq:lambda}) is chosen to
approximate the full radiative transfer models of
\cite[Levermore \& Pomraning (1981)]{levermore1981}.

\section{Orbital Period and Opacity Variations}
In this section I present simulations solving Equations
(\ref{eq:continuity})-(\ref{eq:thermalenergy}) in a planet similar to
HD 209458b, but for a variety of opacities and orbital
periods. Opacity is a crucial factor in determining the efficiency of
energy transport through the atmosphere. Therefore, I present separate
results utilizing both the higher interstellar opacities
(\cite[Pollack et al.  (1985)]{pollack1985} for lower temperatures
coupled with \cite[Alexander \& Ferguson (1994)]{alexander1994} for
higher temperatures) and also lower line opacities of \cite[Freedman
et al. (2008)]{freedman2008}. Both sets of opacities are temperature
and pressure dependent. Although it is now thought that the opacities
of \cite[Freedman et al. (2008)]{freedman2008} more closely represent
the actual planetary opacities, it is clear that there is great
diversity among the observed properties. Opacity may likely be a major
factor in this diversity, so understanding how the dynamics and energy
transfer change with varying opacity is quite important.

\begin{figure}
\centering
\resizebox{6.5cm}{!}{\includegraphics{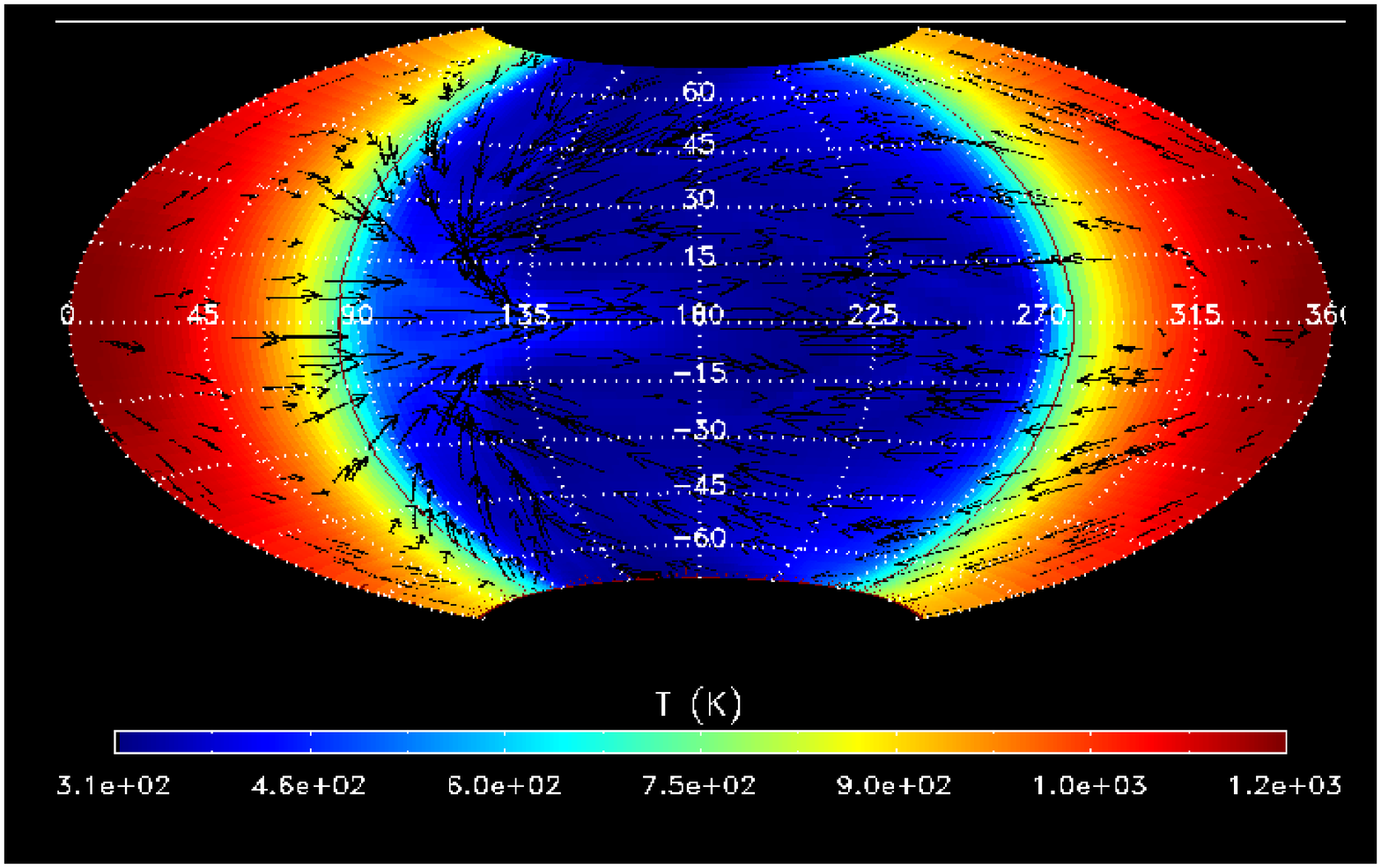} }
\resizebox{6.5cm}{!}{\includegraphics{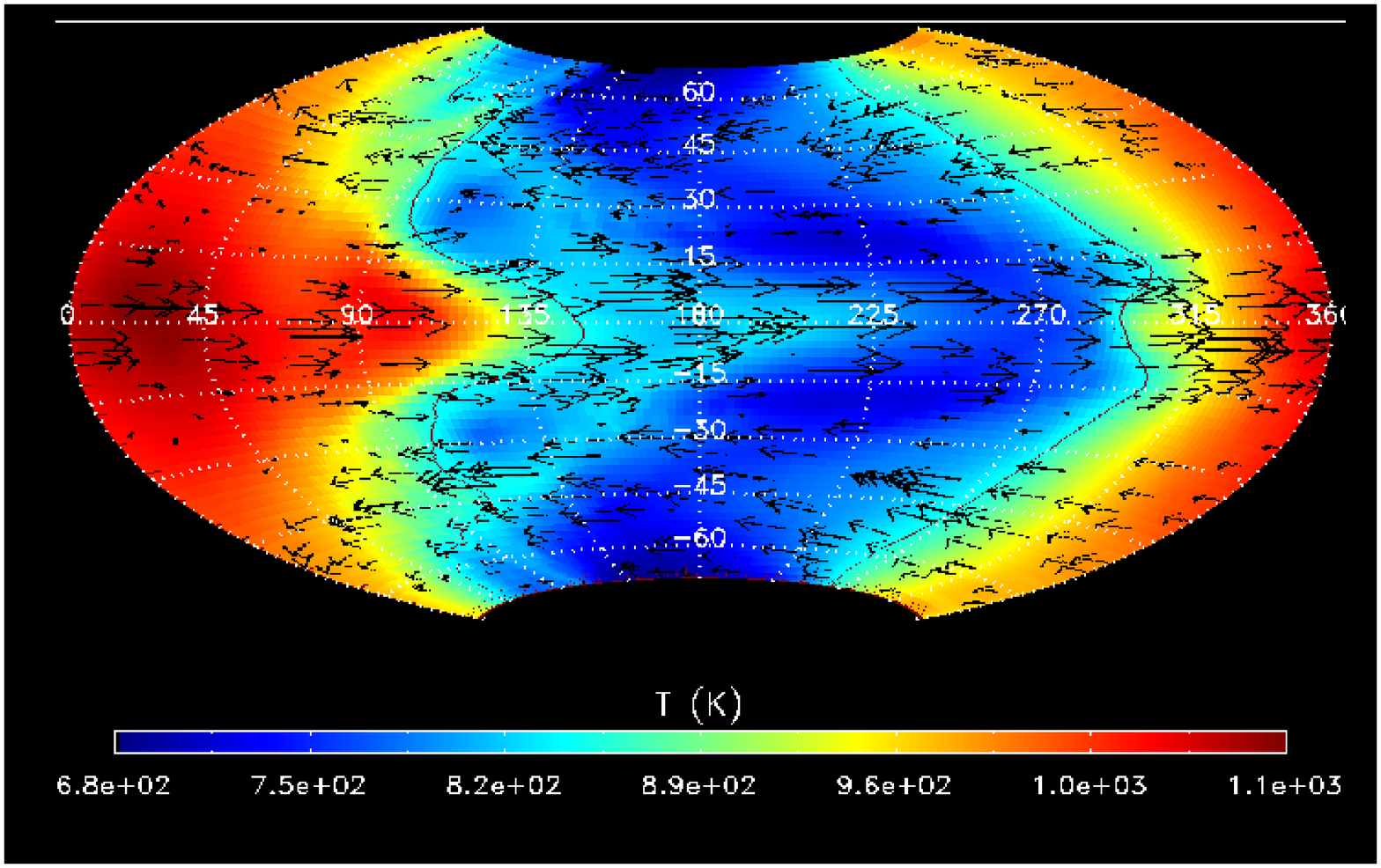}}
\caption[]{Temperature variations at the photosphere of irradiated
planets synchronously rotating at orbital periods of 3-days. The left
panel shows results using large interstellar opacities, while the
right panel shows results using lower \cite[Freedman et
al. (2008)]{freedman2008} opacities. The subsolar point is located at
zero degrees. Photospheric velocities are shown with the over-plotted
arrows.}
\label{fig:Tphoto1}
\end{figure}

In Figure (\ref{fig:Tphoto1}) I show the effect of opacity on the
temperature at the photosphere of a synchronously rotating planet at
an orbital distance of $0.04$ AU. As is clear from Figure
(\ref{fig:Tphoto1}), the higher the opacity the lower the night-side
temperature. Night-side temperatures in a planet with high
interstellar opacities are around $300$ K, while the night-side of a
planet with lower opacity is around $700$ K. This difference can be
easily understood by considering that the depth at which radiation is
deposited changes with opacity. For large opacities the stellar energy
is deposited high in the atmosphere where cooling timescales are
relatively short. The flow is not able to effectively advect energy
around the planet before it cools. For the lower opacity model, the
stellar energy is deposited deeper in the atmosphere where cooling
timescales become comparable to the crossing timescale. The efficiency
of energy advection is also evident in the lobes that extend onto the
night-side at both the equator (in the eastward direction) and the
poles (in the westward direction). These regions correspond to the
largest eastward and westward flow velocities respectively, thus the
shortest crossing timescales. In general the night-side temperature
can be calculated by equating the crossing and radiative timescales
(\cite[Burkert et al. (2005)]{burket2005}). This gives a opacity
dependent temperature that can be expressed as
\begin{equation}
T_n = \left(\frac{4vc_s^2}{3\pi\kappa_d\sigma R_p}\right)^{1/4}.
\end{equation}
This is explored in greater detail in \cite[Dobbs-Dixon \& Lin
(2008)]{dobbsdixon2008}.

\begin{figure}
\centering
\resizebox{6.5cm}{!}{\includegraphics{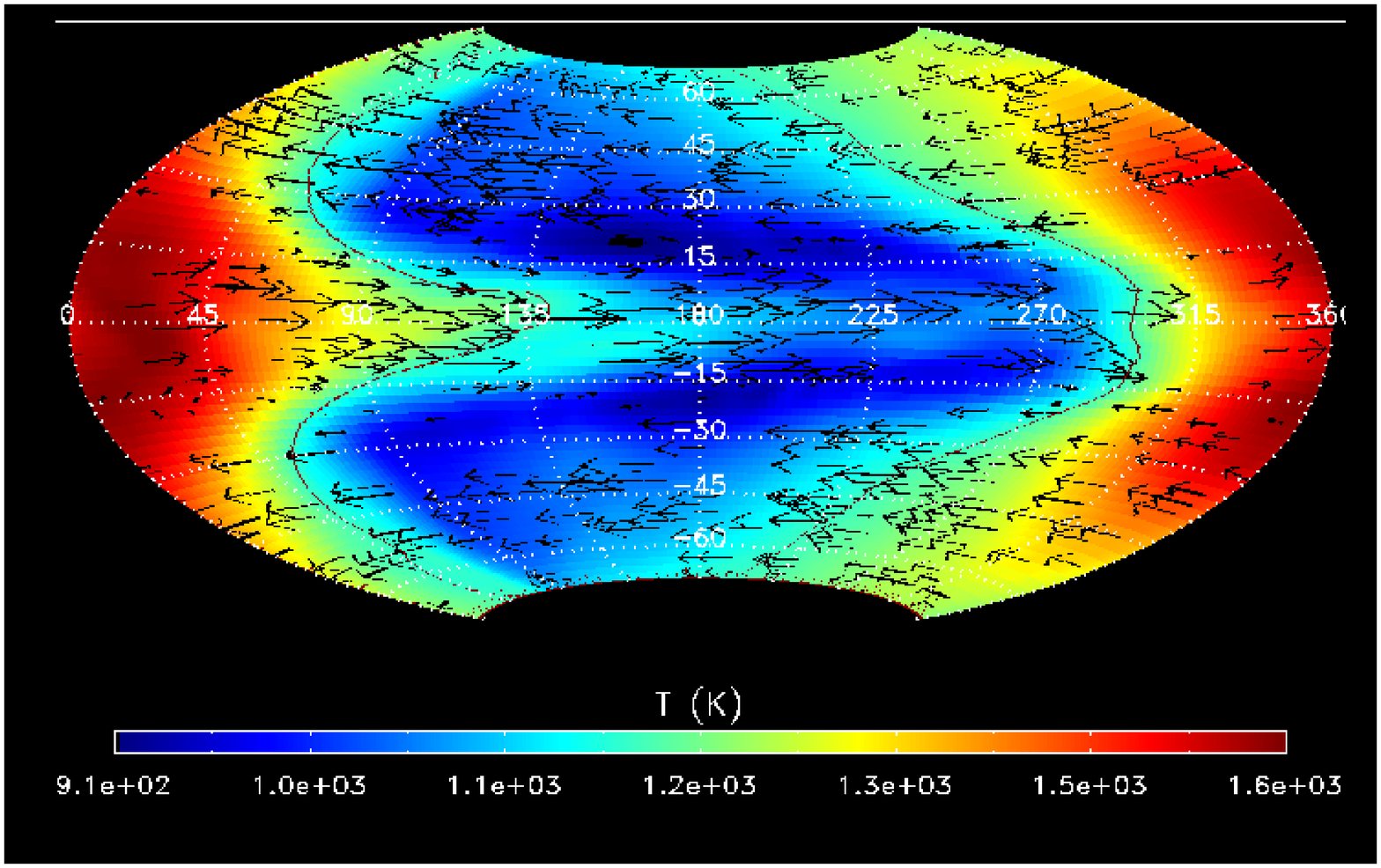} }
\resizebox{6.5cm}{!}{\includegraphics{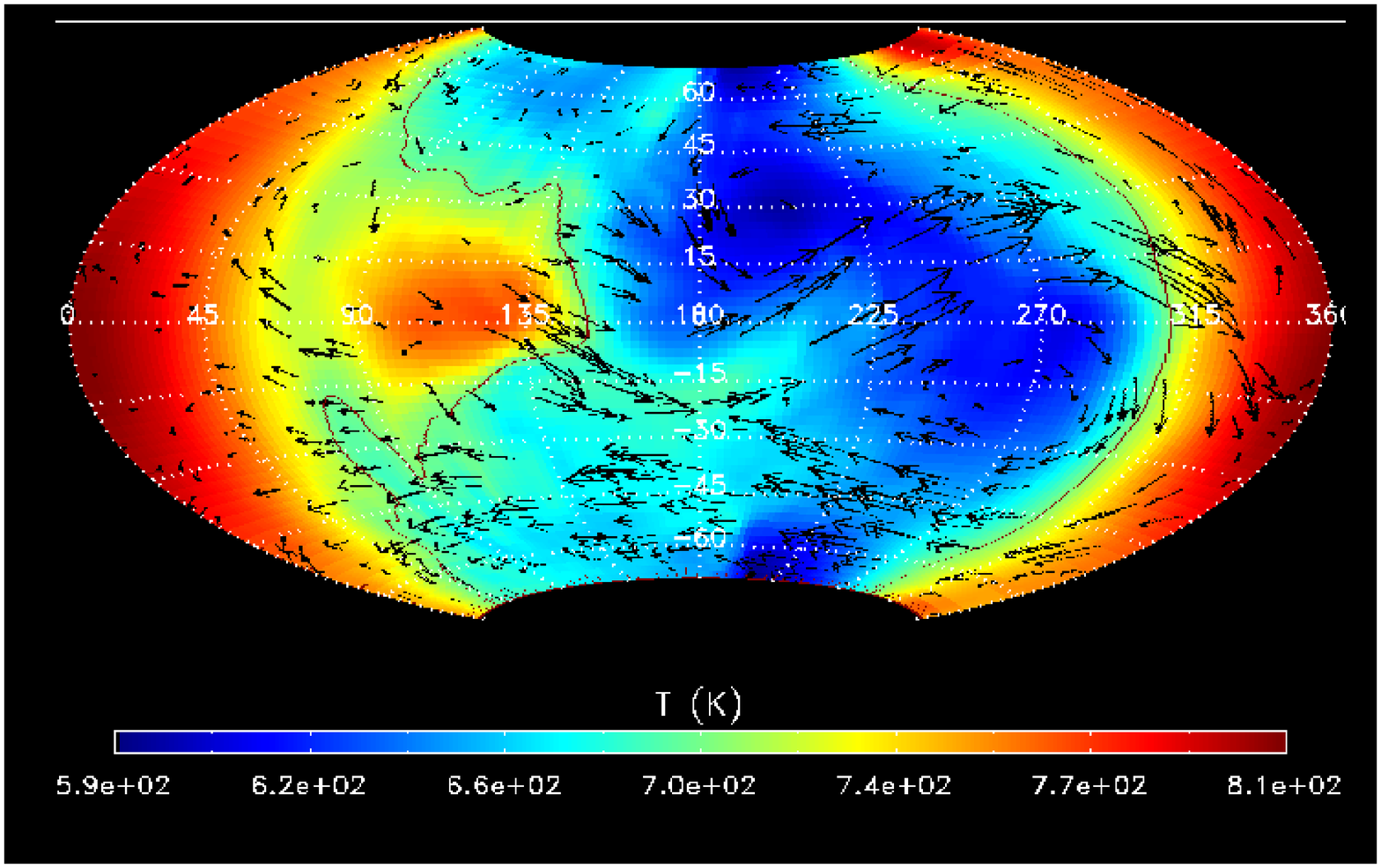}}
\caption[]{Temperature variations at the photosphere of planets with
orbital periods of 1 (left panel) and 7 (right panel) days. Clear
structures associated with stable jets are evident in the 1-day
simulation, while the night-side temperature of the highly variable 7
day simulation is much more uniform.}
\label{fig:rotation}
\end{figure}

Given the rapidly developing pace of the field it is also worthwhile
to examine the atmospheric dynamics of planets for a range orbital
periods. Given the short timescale for synchronization I assume that
the rotation and orbital frequencies are tidally locked in each
model. The problem of non-synchronous planets (whose subsolar points
move across the planet with time) is also an interesting and relevant
question and will be examined in \cite[Dobbs-Dixon et
al. (2008)]{dobbsdixon2008}. In Figure (\ref{fig:rotation}) I show the
results of simulations with orbital and rotation periods of 1 and 7
days, with incident fluxes scaling as $P_{orb}^{-4/3}$. These
simulations were completed using the opacities of \cite[Freedman et
al. (2008)]{freedman2008} and should be compared to the $P_{orb}=3$
day simulation shown on the right side of Figure (\ref{fig:Tphoto1}).

The clearly evident banded structure in the simulations of 1 and 3
days is a direct result of the latitudinal Coriolis force
$-2\Omega_{p}v_{\phi}sin(\theta)$. The formation of the equatorial jet
comes about as eastward moving material feels a force funnelling it
toward the equator, while the oppositely moving westward flow is
pushed toward the poles. Faster rotation results in stronger
funnelling and narrower jets.

The behavior of the simulation with an orbital period of $7$ days is
somewhat different. In general the structure of an eastward
equatorial jet and westward jets at higher and lower latitudes'
persists. However, the day-night temperature difference in this
simulations is smaller. The jet velocity, driven by the temperature
differential, is substantial smaller and the resulting temperature and
velocity structure on the night-side is quite variable.

\section{Decoupled Energy Equations: Temperature Inversions}
As spectral observations become available for an increasing number of
transiting planets, the presence of temperature inversions in several
systems have become evident ([eg.]\cite[Burrows et
al. (2007)]{burrows2007}). Although suggested by \cite[Hubeny et
al. (2003)]{hubeny2003}, prior to these observations this notion had
not been pursued. The physics behind this inversion can be understood
in terms of the different opacities for optical and infrared
photons. The opacity of the impinging stellar energy (optical) can be
significantly different then the opacity of the reprocessed infrared
radiation from the planet. Although a one-dimensional wavelength
dependent radiative transfer model implicitly accounts for these
effects by modelling the impinging stellar spectrum, full radiative
hydrodynamical models must use average grey opacities due to
computational considerations.  In order to encapsulate this physics
within a full hydrodynamical model I have calculated separate grey
opacities for the incident stellar light and the reprocessed
light. Both are still functions of the local temperature and pressure,
but the stellar opacities are weighted toward optical wavelengths.

Furthermore, to allow for the inversion I have de-coupled the thermal
and radiative energy components. Each can now evolve independently,
though they are linked by the appropriate emission and absorption
terms. Although computationally more intensive, this is a significant
improvement to previously employed models. The thermal energy equation
is given by
\begin{equation}
\left[ \pdt{\epsilon} + ({\bf u}\cdot\nabla) \epsilon \right] = - P \,
\nabla \cdot {\bf u} - \rho
\kapp\left[\bbody - c\er \right] +
\rho\kapa F_{\star} e^{-\tau_\star},
\label{eq:thermalenergy2}
\end{equation}
while the independent radiation energy equation is given by
\begin{equation}
\pdt{\er} + \nabla \cdot {\bf F} = \rho
\kapp\left[\bbody-c\er\right].
\label{eq:radenergy}
\end{equation}
As before, $\epsilon=c_{v} \rho T$ is the internal energy density, $T$
is the local gas temperature, and $c_v$ is the specific heat. In
addition, $\bbody=4\sigma T^4$, $\kapa$ is the absorption opacity
\emph{averaged with the stellar temperature}, and $F_{\star}$ is the
impinging stellar flux. The term proportional to $\bbody-c\er,$
represents the exchange of energy between the thermal and radiative
components through the emission and absorption of the lower energy
photons. The $\rho\kapa F_{\star} e^{-\tau_{\star}}$ term represents
the higher energy stellar photons absorbed by the gas. Given the
relatively low local temperatures, I do not consider the energy
re-emitted into the higher energy waveband. The radiative flux is
treated using the flux-limited diffusion method described in Equation
(\ref{eq:flux}). Equations (\ref{eq:thermalenergy2}) and
(\ref{eq:radenergy}) contain three separate opacities: the optical
absorption opacity $\kappa_A$, the infrared Planck opacity $\kappa_P$,
and the infrared Rosseland opacity $\kappa_R$ that appears through the
flux term.

\begin{figure}
\centering
\resizebox{6.5cm}{!}{\includegraphics{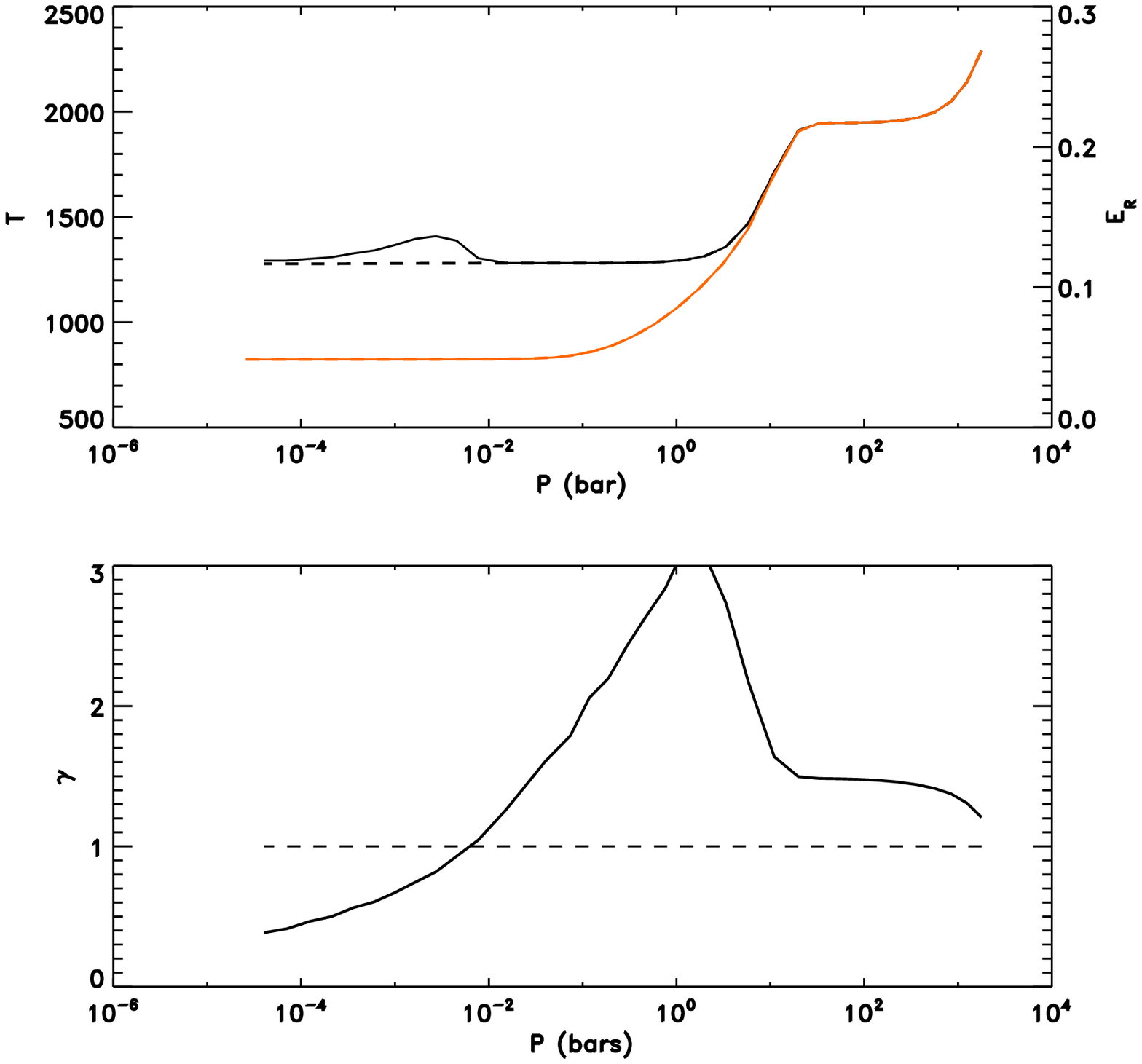}}
\resizebox{6.5cm}{!}{\includegraphics{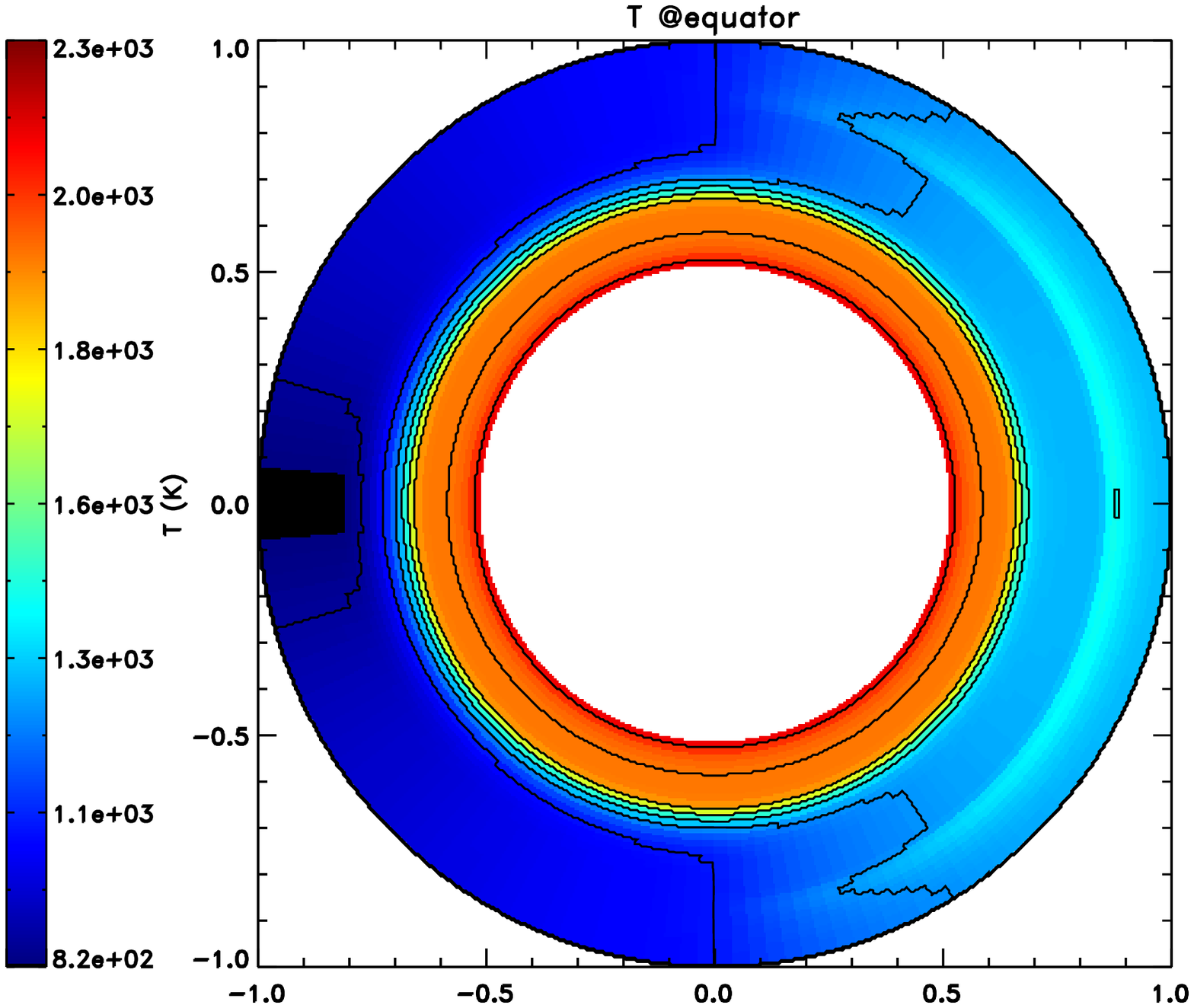}}
\caption[]{The temperature in HD 209458b concentrating only on the
radiative solution to Equations (\ref{eq:thermalenergy2}) and
(\ref{eq:radenergy}). The left hand panel shows the
temperature-pressure profile at the subsolar point. The inversion in T
(solid line) is quite clear. The dashed line shows the decoupled
radiative energy $\er$. The orange line shows the temperature profile
at the anti-stellar point. The lower left panel shows the behavior of
$\gamma$ with pressure. The right panel shows the entire equatorial
slice of the same simulation with the star to the right.}
\label{fig:PT_T_rad}
\end{figure}

The effect of decoupling the energy equations and including 3 separate
opacities can best be understood by considering the radiative behavior
of Equations (\ref{eq:thermalenergy2}) and (\ref{eq:radenergy}), {\it
ie.} in the absence of dynamics. If we express the ratio of the
stellar absorption opacity to the local opacity as (\cite[Hubeny et
al. (2003)]{hubeny2003})
\begin{equation}
\gamma^4 \equiv \frac{\kapa}{\kapp},
\end{equation}
then the steady-state, radial temperature profile can be expressed as
\begin{equation}
T^4(r) = \frac{\gamma^4(r) T_0^4 e^{-\tau_{\star}}}{4}
  -\frac{\er (r)}{a}.
\end{equation}
The temperature in the upper atmosphere, near the stellar photosphere
($\tau_{\star}\sim1$) will scale as $\gamma T_0$, where $T_0$ is the
equilibrium stellar temperature at the planetary semi-major axis. The
presence of an inversions thus becomes dependent on the radial
structure of $\gamma$. In Figure (\ref{fig:PT_T_rad}) I show the
solution to Equations (\ref{eq:thermalenergy2}) and
(\ref{eq:radenergy}) in the absence of dynamics. As expected, the
inversion is high in the atmosphere, where $\gamma$ is around or above
unity and $\tau_{\star}$ is small. Figure (\ref{fig:PT_T_rad}) clearly
demonstrates how the thermal and radiative energies become de-coupled,
requiring solving the equations separately. Full dynamical simulations
are not presented here, but will be in explored in detail in
\cite[Dobbs-Dixon \& Lin (2008)]{dobbsdixon2008}.

\section{Discussion}
In this proceeding I have reviewed results from full three-dimensional
radiative hydrodynamical simulations of the atmospheres of close-in
planets. Simulations for a range of atmospheric opacities and
semi-major axis were performed. The results indicate that both stellar
proximity and opacity are both crucial factors in determining the
behavior of planetary atmospheres. Increasing planetary opacities
yields less efficient advection and colder night-side
temperatures. Furthermore, planets situated at 1 and 3 day orbital
periods exhibit a fairly stable banded jet structure, with jet width
increasing with rotation period. On the other-hand, dynamics of
planets with 7 day orbital periods can be much more variable, with
wandering jets and the formation of vortices on the night-side. 

The decoupling of the thermal and radiative energies, as well as the
introduction of multiple opacities relevant for impinging optical
stellar photons vrs local re-radiated infrared photons allows us to
explore in greater detail temperature structure near the top of the
atmosphere. Radiative solutions presented in Section (4) produce
temperature inversions similar to those observed and demonstrate the
feasibility of simultaneously studying these radiative features and the
resulting hydrodynamics. The existence of an inversion depends
critically on the ratio of the optical to infrared opacities and the
behavior of this ratio with radius.

Given the importance of both optical and infrared opacities it is
fruitful to consider how we expect atmospheric properties to vary as
we change these two quantities independently. \cite[Fortney et
al. (2008)]{fortney2008} suggested a classification planets based on
the presence or absence of TiVO in the upper atmosphere. Although the
presence of TiVO in the upper atmosphere is controversial, it would
provide the necessary absorption opacity to cause an
inversion. Fortney's classifications of pM and pL classes correspond
to those with and without detected inversions. Here extend this
classification in terms of the individual absorption ($\kapa$) and
Planck ($\kapp$)opacities explored in this proceeding.

\begin{figure}
\centering
\resizebox{8.0cm}{!}{\includegraphics{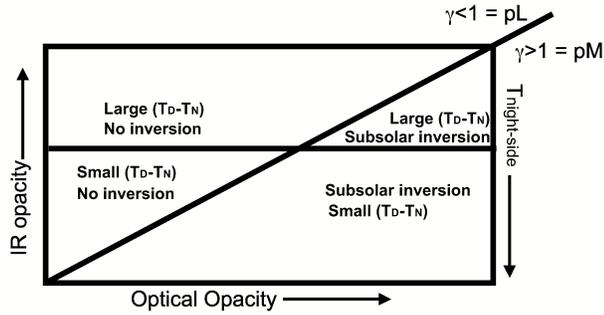}}
\caption[]{The gross behavior of irradiated atmospheres as you change
the absorption and local Planck opacities. Increasing optical
absorption opacity ($\kapa$) is given in the horizontal axis, while
the vertical axis gives the local opacity to re-radiated photons
($\kapp$). The diagonal line demarcates the division between planets
with an inversion ($\gamma>1$) and those without ($\gamma<1$), where
$\gamma^4=\kapa/\kapp$. Phase measurements coupled to inversion
determinations will uniquely place you in a given quadrant.}
\label{fig:diagram}
\end{figure}

As explained in Section (3), the night-side temperature is governed by
the efficiency of advection, which in turn is determined by the
cooling timescale where the incident radiation is deposited. The
cooling timescale is directly related to the local Planck opacity
$\kapp$; large $\kapp$ will result in a large day-night temperature
differential. Such a distinction is observable from current and
ongoing phase monitoring programs. On the other hand, the absorption
opacity for the incident stellar light $\kapa$ will play an important
role in the presence or absence of a temperature inversion. As
described in Section (4) the ratio $\gamma=( \frac{\kapa}{\kapp} )
^{1/4}$ will determine the temperature structure in the upper
atmosphere. If the radial structure of $\gamma e^{-\tau_{\star}}$ is
such that it peaks in the upper atmosphere there will be an
inversion. In these terms, the presence of TiVO (a pL planet) would
lead to large $\kapa$, but an inversion would occur only if $\kapp$
does not also increase. In Figure (\ref{fig:diagram}) I have unified
these two concepts onto a single diagram. A measurement of both the
magnitude of day-night temperature differential and a determination of
the presence or absence of an inversion will uniquely tell you both
the optical and infrared opacities and place the planet in one of the
quadrants. Such a determination will ultimately tell us much about the
formation and evolution of these planets.

\end{document}